\documentclass[aps,pra,preprint,floatfix,amsmath,showpacs]{revtex4}
\usepackage{amsmath}
\usepackage{amssymb}
\usepackage{graphicx}
\usepackage{subfigure}

\begin{document}

\title{The Ginzburg-Landau model of Bose-Einstein condensation of magnons}
\author{B. A. Malomed$^{1}$, O. Dzyapko$^{2}$, V. E. Demidov$^{2}$, S. O.
Demokritov$^{2}$}
\affiliation{$^{1}$Department of Physical Electronics, School of Electrical Engineering,
Tel Aviv University, Tel Aviv 69978, Israel\\
$^{2}$Institute for Applied Physics, University of M\"{u}nster,
48149 M\"{u}nster, Germany}

\begin{abstract}
We introduce a system of phenomenological equations for
Bose-Einstein condensates of magnons in the one-dimensional
setting. The nonlinearly coupled equations, written for amplitudes
of the right-and left-traveling waves, combine basic features of
the Gross-Pitaevskii and complex Ginzburg-Landau models. They
include localized source terms, to represent the microwave
magnon-pumping field. With the source represented by the $\delta
$-functions, we find analytical solutions for symmetric localized
states of the magnon condensates. We also predict the existence of
asymmetric states with unequal amplitudes of the two components.
Numerical simulations demonstrate that all analytically found
solutions are stable. With the $\delta $-function terms replaced
by broader sources, the simulations reveal a transition from the
single-peak stationary symmetric states to multi-peak ones,
generated by the modulational instability of extended
nonlinear-wave patterns. In the simulations, symmetric initial
conditions always converge to symmetric stationary patterns. On
the other hand, asymmetric inputs may generate nonstationary
asymmetric localized solutions, in the form of traveling or
standing waves. Comparison with experimental results demonstrates
that the phenomenological equations provide for a reasonably good
model for the description of the spatiotemporal dynamics of magnon
condensates.
\end{abstract}

\maketitle

\section{Introduction}

The experimental observation of the Bose-Einstein condensation (BEC) in
dilute gases of alkali atoms \cite{BEC} has been a milestone in the
development of atomic and condensed-matter physics, demonstrating the
reality of this state of matter, and providing a unique testbed for studying
numerous macroscopic quantum effects \cite{review}. As is well known, the
transition to the condensation in atomic gases occurs at extremely low
temperatures, $\sim 10^{-7}$ K, the number of atoms in the condensate
usually being quite small, $\lesssim 10^{4}$.

These achievements were followed by the creation of BEC in gases
of bosonic quasi-particles representing fundamental excitations in
solid-state media, including excitons \cite{exciton}, polaritons
\cite{polariton}, triplons \cite{triplon} and magnons (spin waves
in ferromagnets) \cite{magnon}. Similar to the BEC of atomic
gases,\ the condensation of quasi-particles takes place when their
density exceeds a certain critical value, which increases with the
temperature of the system. However, the advantage of studying BEC
in solids is that the density may be increased by external pumping
fields: laser or microwave radiation, in the case of excitons and
polaritons, or magnons, respectively. Therefore, the BEC
transition in solid-state media can be reached at much higher
temperatures than in the atomic gases.

Unique properties are featured by the BEC of magnons in tangentially
magnetized yttrium-iron-garnet (YIG) films. In particular, the condensation
of magnons occurs at room temperatures \cite{magnon,incoherent}. Moreover,
while, in the first experiments, monochromatic pumping fields were used, it
has been recently reported that the BEC of magnons can be achieved by means
of an incoherent microwave pumping \cite{incoherent}.

Despite the achievements in the experimental studies of BEC of
magnons, development of the corresponding theory is just starting
\cite{Loktev1,Tupitsin,threshold2,threshold3}. Thermodynamics of
the quasi-equilibrium magnon gas and the stability of the
condensate were considered in Refs. \cite{Loktev1} and
\cite{Tupitsin} respectively. The formation of the coherent magnon
state (the BEC proper), accompanied by the emergence of
macroscopic dynamic magnetization, is described in Ref.
\cite{threshold2}. The theory is capable of explaining, in a
quantitative form, the observed intensities of the Brillouin light
scattering and microwave signals generated by the condensate. The
presented work addresses the dynamics of the magnon condensate,
first experimental studies of which were recently reported in Ref.
\cite{threshold}.

The magnon BEC is fundamentally different from its counterparts in all other
systems. The condensation of magnons occurs at two separate points in the
space of the magnon frequency and wavenumber, $(\omega _{0},\pm k_{0})$,
hence, for the theoretical description of the condensate one needs to
introduce wave functions with two components corresponding to these points.
One should take into account interactions between these two components,
which may result in effects which are not observed in BEC of other
(quasi-)particles.

In accordance with the above discussion, different types of models
are necessary for the description of BEC in atomic gases and
solid-state media have to be used. In the former case, a widely
adopted theoretical approach is based on the Gross-Pitaevskii
equation (GPE) for the single-atom wave function \cite{review}. In
the mean-field approximation, the GPE takes into account
collisions between atoms through the cubic term. In fact, the GPE
provides for a very accurate model for the practically relevant
case of dilute ultra-cold gases. An inherent property of the GPE
is that the norm of the wave function is a dynamical invariant,
which complies with the obvious condition of the conservation of
the number of atoms in the gas. Equations that may serve as a
mean-field model of the condensate of quasi-particles should be
different, as the quasi-particles (in particular, magnons) may be
created by the pump field and lost through linear and nonlinear
dissipation. Therefore, the respective equation is expected to be
of the complex-Ginzburg-Landau (CGL)\ type \cite{CGL}, combining
conservative and dissipative terms. However, a difference from the
usual form of the CGL equation is that, in the present case, it
must include a source term representing the localized pump (in
this respect, the equation may be somewhat similar to those
describing nonlinear optical cavities pumped by external laser
beams \cite{cavity}). Nevertheless, this difference is not
absolute, as ultra-cold atoms have an effective finite lifetime,
due to heating effects and three-body interactions. As discussed
in Refs. \cite{SnokeLifetime}, the ratio between the lifetime of
the (quasi-)particles and their thermalization time is decisive
for this issue. For magnons this factor takes values in the range
of $5-10$, whereas for ultra-cold alkali atoms it may reach two
orders of magnitude.

The objective of the present work is to put forward a semi-phenomenological
system of coupled equations for the wave functions of the magnon condensate
generated by the local microwave source. Comparing predictions of the model
with experimental observations of the dynamics reported in \cite{threshold},
we conclude that the equations correctly reproduce main features of the
spatio-temporal dynamics (pattern formation in the condensate). The
equations are introduced in Section II, and some exact analytical solutions
for localized states, predicted by the model, are reported in Section III.
This is followed by the presentation of numerical solutions in Section IV
and, eventually by the comparison with the experiment in Section V. The
paper is concluded by Section VI.

\section{The model}

As mentioned above, the onset of the condensation in the magnon
gas reveals itself in the emergence of the macroscopic dynamical
magnetization in the sample. It should be emphasized that only
condensed magnons, with the above-mentioned values of the
frequency and wavenumbers, $(\omega _{0},\pm k_{0})$, contribute
to this magnetization, whereas all other magnons merely decrease
the absolute value of the static magnetization. Following the
envelope (slowly-varying-amplitude) approximation, widely adopted
for the analysis of
nonlinear spin waves in ferromagnetic films \cite{NLSW1,NLSW2,ZvezdinPopkov}%
, as well for the description of nonlinear waves in many other settings, we
introduce the order parameter (the full wave function of magnons) as%
\begin{equation}
\Psi \left( z,t\right) =\Psi _{+}(z,t)e^{ik_{0}z}+\Psi _{-}(z,t)e^{-ik_{0}z},
\label{Psi}
\end{equation}%
where absolute values of amplitudes $\Psi _{\pm }$ represent dynamical
macroscopic magnetizations created by the corresponding condensate
components, $m_{\pm }$, normalized to the absolute value of the static
magnetization ($M_{0}$): $\left\vert \Psi _{\pm }\right\vert ^{2}=\left\vert
m_{\pm }\right\vert ^{2}/\left( 2M_{0}^{2}\right) $, while phases of  $\Psi
_{\pm }$ account for the coherence of the two components of the condensate.
In Eq. (\ref{Psi}), it is assumed that the magnetic field is aligned with
axis $z$, the microwave antenna is oriented perpendicular to it, and $\pm
k_{0}$ are large wavenumbers ($k_{0}\simeq 3\times 10^{4}$ cm$^{-1}$ in YIG
films) at which the transition into the BEC state takes place. The remaining
dependence on $z$ in $\Psi _{\pm }(z,t)$ is assumed to be slow in comparison
with the rapidly oscillating carrier waves, $\exp \left( \pm ik_{0}z\right) $%
, therefore the interaction between $\Psi _{+}$ and $\Psi _{-}$ is
incoherent [see Eqs. (\ref{Psi+}) and (\ref{Psi-}) below]. The slow
dependence arises due local variations of the pumping which excites magnons.

In Refs. \cite{NLSW1,NLSW2,ZvezdinPopkov} it was shown that, in the case of
single-component magnon wave function, the evolution of the envelope is
described by the nonlinear Schr\"{o}dinger equation. In a similar way, for
two components $\Psi _{\pm }$ one arrives at a system of phenomenological
equations of the GPE/CGL type, which include the above-mentioned source
terms, written below as $-f\delta (z)$ to allow analytical treatment of the
problem (numerical results will be reported below for a realistic broad
shape of the source). In the scaled form, the equations are:%
\begin{gather}
i\left( \Psi _{+}\right) _{t}+\delta \mu _{0}\cdot \Psi _{+}+\frac{1}{2}%
\left( \Psi _{+}\right) _{zz}+i\eta \Psi _{+}  \notag \\
+\left( \left\vert \Psi _{+}\right\vert ^{2}+\sigma _{1}\left\vert \Psi
_{-}\right\vert ^{2}\right) \Psi _{+}+i\tau \left( \left\vert \Psi
_{+}\right\vert ^{2}+\sigma _{2}\left\vert \Psi _{-}\right\vert ^{2}\right)
\Psi _{+}=-f~\delta (z).  \label{Psi+}
\end{gather}%
\begin{gather}
i\left( \Psi _{-}\right) _{t}+\delta \mu _{0}\cdot \Psi _{-}+\frac{1}{2}%
\left( \Psi _{-}\right) _{zz}+i\eta \Psi _{-}  \notag \\
+\left( \left\vert \Psi _{-}\right\vert ^{2}+\sigma _{1}\left\vert \Psi
_{+}\right\vert ^{2}\right) \Psi _{-}+i\tau \left( \left\vert \Psi
_{-}\right\vert ^{2}+\sigma _{2}\left\vert \Psi _{+}\right\vert ^{2}\right)
\Psi _{-}=-f~\delta (z).  \label{Psi-}
\end{gather}%
Here, $\delta \mu _{0}$ is a possible shift of the chemical potential of
magnons in the condensate with respect to that of uncondensed magnons. We
also introduce phenomenological damping, represented by linear and cubic
terms in Eqs. (\ref{Psi+}) and (\ref{Psi-}), with coefficients $~\eta $ and $%
\tau $, respectively. The same nonlinear-damping term was introduced
in Ref. \cite{NLDamping}. Further, $\sigma _{1}$, $\sigma _{2}$ are
the XPM/SPM (cross/self-phase-modulation) ratios for the
conservative and dissipative nonlinear parts of the equations. The
sign in front of the SPM and XPM nonlinearities in the conservative
part corresponds to the case of attractive magnon-magnon
interaction, which was realized in tangentially magnetized YIG films
in experimental works \cite{ZvezdinPopkov,Tupitsin}.

It is relevant to mention that equations similar to Eqs. Eqs.
(\ref{Psi+}) and (\ref{Psi-}) were derived by Lvov for the
description of spatially inhomogeneous parametric excitation of
magnons \cite{Lvov} (in that case, two groups of magnons with
wavenumbers $\pm k$ were also excited). In fact,
the applicability of amplitude equations of this general (\textit{%
Ginzburg-Landau}) type to the description of condensates is a universal fact
(at least, at the phenomenological level), even in the case of strong
interactions between quasi-particles \cite{LP}.

The notation adopted in the equations implies that $t$ is
normalized to the characteristic time of the nonlinear
interactions of magnons, $\tau _{0}\simeq 3.6$ ns, which, as said
above, is much shorter than the lifetime of a magnon ($T_{m}\simeq
250$ ns in YIG \cite{Coherence}), hence $\eta \equiv \tau
_{0}/\left( 2T_{m}\right) $ is considered as a small parameter.
Spatial coordinate $z$ is normalized to characteristic length
$z_{0}$, which is defined by the ratio of the nonlinear
interaction strength to the
dispersion coefficient ($z_{0}\simeq 0.5$ $\mathrm{\mu }$m$.$ in YIG \cite%
{ZvezdinPopkov}). Unlike other coefficients, which may be identified from
empirical data, $\delta \mu _{0}$ should be treated as a phenomenological
parameter, that may be found from adjusting theoretical predictions to
experimental observations.

The strength of the source, $f$, may be set to be real and positive, by
definition. As mentioned above, function $\delta (z)$ in Eqs. (\ref{Psi+}), (%
\ref{Psi-}) is the Dirac's $\delta $-function if we aim to find analytical
solutions (see below), or a regularized counterpart of the $\delta $%
-function for numerical solutions. In the latter case, it is taken as%
\begin{equation}
\delta (z)\rightarrow \tilde{\delta}(z)\equiv \frac{1}{\sqrt{\pi }\Delta }%
\exp \left( -\frac{z^{2}}{\Delta ^{2}}\right) ,  \label{tilde}
\end{equation}%
$\Delta $ being the regularization parameter.

Note that Eqs. (\ref{Psi+}), (\ref{Psi-}) are somewhat similar to the
coupled CGL equations for counterpropagating waves in the binary-fluid
thermal convection \cite{Cross}. However, that model includes finite group
velocities, but not the source terms. Equations (\ref{Psi+}), (\ref{Psi-})
do not contain group-velocity terms [which might be, generally speaking, $%
\pm ic\left( \Psi _{\pm }\right) _{z}$], because the group velocity of
magnons vanishes at the condensation points, $k=\pm k_{0}$.

In the case of the $\delta $-function, \emph{symmetric} stationary solutions
to Eq. (\ref{Psi}) may be sought for as $\Psi _{+}(z,t)=\Psi _{-}(z)=\Psi (z)
$, where the single complex function, $\Psi (z)$, satisfies an ordinary
differential equation,%
\begin{equation}
\left( \delta \mu _{0}+i\eta \right) \Psi +\frac{1}{2}\frac{d^{2}\Psi }{%
dz^{2}}+\left[ \left( 1+\sigma _{1}\right) +i\tau \left( 1+\sigma
_{2}\right) \right] |\Psi |^{2}\Psi =0,  \label{ODE}
\end{equation}%
at $~z\neq 0$, supplemented by the boundary condition (b.c.) at $z=0$, which
is generated by the integration of Eqs. (\ref{Psi+}) and (\ref{Psi-}) in an
infinitesimal vicinity of $z=0$:%
\begin{equation}
\Psi ^{\prime }|_{z=+0}-\Psi ^{\prime }|_{z=-0}=-2f.  \label{bc}
\end{equation}%
The meaning of this b.c. is that function $\Psi (z)$ must be continuous at $%
z=0$, but its derivative makes a jump at the position of the source. This
approach to searching for stationary solutions is similar to that which was
recently developed for a CGL equation (with an intrinsic gain, but without
source terms) driven by the delta-like gain \cite{HK}.

Concluding this Section, we emphasize that the source term introduced in the
above equations is not direct proportional to the microwave field used in
the experiment for the parametric pumping of magnons. In fact, the
parametric pumping creates primary magnons away from the condensation points
in the phase space. The BEC condensate then forms as a result of the
thermalization of the primary magnons. Thus, the proposed model does not aim
to explicitly describe the spatial spreading of the magnons during the
thermalization process.

\section{Analytical results}

\subsection{Symmetric solutions}

Following the approach of Ref. \cite{HK}, where particular exact solutions
were found in the framework of the above-mentioned CGL equation with the
delta-like linear gain, we look for a localized solution to Eq. (\ref{ODE})
in the following form:
\begin{equation}
\Psi (z)=\frac{Ae^{i\chi }}{\left[ \sinh \left( \kappa \left( |z|+\xi
\right) \right) \right] ^{1-i\nu }},  \label{sinh}
\end{equation}%
with real constants $A,\chi ,\kappa ,\nu $ and $\xi $. In particular, $\xi $
must be positive (otherwise, the expression gives rise to a singularity at $%
\left\vert z\right\vert =-\xi $). Actually, this \emph{ansatz} was suggested
by a formal singular solution to the cubic CGL equation with constant
coefficients, that, in turn, is a counterpart to the so-called
Pereira-Stenflo soliton, with $\sinh $ replaced by $\cosh $ \cite{Lennart}.
The latter is a non-singular but unstable exact solution, which is valid in
the case of $\eta <0$ (a uniform linear gain, instead of the loss).
Solutions of the Pereira-Stenflo type may be made stable as exact solutions
to coupled CGL equations one of which is cubic, and the other one linear
\cite{Javid}.

The substitution of ansatz (\ref{sinh}) in Eqs. (\ref{ODE}) leads to the
following equations:%
\begin{eqnarray}
\kappa ^{2}\left( 2-\nu ^{2}-3i\nu \right) +2\left[ \left( 1+\sigma
_{1}\right) +i\tau \left( 1+\sigma _{2}\right) \right] A^{2} &=&0,  \label{a}
\\
\kappa ^{2}\left( 1-i\nu \right) ^{2}+2\left( \mu _{0}+i\eta \right)  &=&0,
\label{b}
\end{eqnarray}%
Further, inserting ansatz (\ref{sinh}) into b.c. (\ref{bc}), we arrive at an
additional relation,%
\begin{equation}
\frac{\cosh \left( \kappa \xi \right) }{\left[ \sinh \left( \kappa \xi
\right) \right] ^{2-i\nu }}=\frac{fe^{-i\chi }}{A\kappa \left( 1-i\nu
\right) }~.  \label{b.c.}
\end{equation}%
Equations (\ref{a}) and (\ref{b}) can be solved directly [actually, it is
the single physically relevant solution corresponding to ansatz (\ref{sinh}%
)]:%
\begin{eqnarray}
\nu  &=&\sqrt{2+\frac{9\left( 1+\sigma _{1}\right) ^{2}}{4\left( 1+\sigma
_{2}^{2}\right) \tau ^{2}}}+\frac{3\left( 1+\sigma _{1}\right) }{2\left(
1+\sigma _{2}\right) \tau }~,~  \label{nu} \\
\kappa ^{2} &=&\frac{\eta }{\nu }\equiv \frac{\eta }{2}\left[ \sqrt{2+\frac{%
\left( 1+\sigma _{1}\right) ^{2}}{\left( 1+\sigma _{2}\right) ^{2}\tau ^{2}}}%
-\frac{\left( 1+\sigma _{1}\right) }{\left( 1+\sigma _{2}\right) \tau }%
\right] ,  \label{kappa^2}
\end{eqnarray}%
\begin{equation}
A^{2}=\frac{3\eta }{2\tau \left( 1+\sigma _{2}\right) },  \label{A^2}
\end{equation}%
\begin{equation}
\delta \mu _{0}=\frac{1}{2}\left( \nu ^{2}-1\right) \kappa ^{2}.  \label{mu0}
\end{equation}

It is worthy to emphasize the meaning of Eq. (\ref{mu0}): the solution given
by ansatz (\ref{sinh}) exists solely for the particular value of $\delta \mu
_{0}$ given by this expression. For other values of $\delta \mu _{0}$,
localized solutions are also expected to exist (see numerical results
reported below), but they cannot be found in the exact form corresponding to
the present ansatz.

Once constants $\nu $, $\kappa $ and $A$ are known, as per Eqs. (\ref{nu}) -
(\ref{A^2}), the equation corresponding to b.c. (\ref{b.c.}) can be solved
analytically too, and it also produces a single\emph{\ }solution, which
determines the remaining arbitrary constants, $\xi $ and $\chi $:%
\begin{eqnarray}
\cosh \left( \kappa \xi \right) &=&\sqrt{1+\frac{A^{2}\kappa ^{2}\left(
1+\nu ^{2}\right) }{4f^{2}}}+\frac{A\kappa \sqrt{1+\nu ^{2}}}{2f},
\label{xi} \\
\chi &=&\arctan \left( \nu \right) -\nu \ln \left( \sinh \left( \kappa \xi
\right) \right) .  \label{chi}
\end{eqnarray}%
A noteworthy feature of solution (\ref{xi}) is the absence of a threshold:
any pump strength $f$, even a very small one, supports the solution. In the
limit of $f\rightarrow 0$, Eq. (\ref{xi}) yields $\xi \rightarrow \infty $,
which naturally implies the disappearance of the BEC state when the
microwave source is switched off.

\subsection{Asymmetric solutions}

A challenging issue for the theory is a possibility of the existence of
asymmetric solutions, with $\Psi _{+}(z)\neq \Psi _{-}(z)$. In that case,
the stationary version of Eqs. (\ref{Psi+}) and (\ref{Psi-}) takes the
following form:%
\begin{gather}
\delta \mu _{0}\cdot \Psi _{+}+\frac{1}{2}\left( \Psi _{+}\right)
_{zz}+i\eta \Psi _{+}  \notag \\
+\left( \left\vert \Psi _{+}\right\vert ^{2}+\sigma _{1}\left\vert \Psi
_{-}\right\vert ^{2}\right) \Psi _{+}+i\tau \left( \left\vert \Psi
_{+}\right\vert ^{2}+\sigma _{2}\left\vert \Psi _{-}\right\vert ^{2}\right)
\Psi _{+}=-f~\tilde{\delta}(z).  \label{asymm+}
\end{gather}%
\begin{gather}
\delta \mu _{0}\cdot \Psi _{-}+\frac{1}{2}\left( \Psi _{-}\right)
_{yy}+i\eta \Psi _{-}  \notag \\
+\left( \left\vert \Psi _{-}\right\vert ^{2}+\sigma _{1}\left\vert \Psi
_{+}\right\vert ^{2}\right) \Psi _{-}+i\tau \left( \left\vert \Psi
_{-}\right\vert ^{2}+\sigma _{2}\left\vert \Psi _{+}\right\vert ^{2}\right)
\Psi _{-}=-f~\tilde{\delta}(z).  \label{asymm-}
\end{gather}%
Here, we assume that $\tilde{\delta}(z)$ is not an infinitely
narrow delta-function, but rather a regularized one -- e.g., as
given by Eq. (\ref{tilde}). We will demonstrate below that in this
case an asymmetric solution
appears spatially nonuniformly, i.e., in regions of $z$ where $\tilde{\delta}%
(z)$ exceeds a certain threshold.

The possibility of the existence of asymmetric solutions can be
investigated analytically under the assumption that the linear and
nonlinear dissipative coefficients, $\eta $ and $\tau $, are large
parameters. Strictly speaking, this assumption contradicts the
above-mentioned condition for the applicability of the equations to
the description of the physically relevant situation, $\eta \ll 1.$
Nevertheless, the analysis makes sense, in view of the fundamental
significance the possibility of the symmetry breaking in models of
the present type.

Assuming that $\eta $ and $\tau $ are large, In the lowest
approximation we neglect all conservative terms and replace Eqs.
(\ref{asymm+}) and (\ref{asymm-}) by the following ones, for $\Psi
_{\pm }(z)\equiv i\Phi _{\pm }(z)$, where functions $\Phi _{\pm
}(z)$ are real:
\begin{equation}
\eta \Phi _{+}+\tau \left( \Phi _{+}^{2}+\sigma _{2}\Phi _{-}^{2}\right)
\Phi _{+}=f~\tilde{\delta}(z).  \label{Phi+}
\end{equation}%
\begin{equation}
\eta \Phi _{-}+\tau \left( \Phi _{-}^{2}+\sigma _{2}\Phi _{+}^{2}\right)
\Phi _{-}=f~\tilde{\delta}(z).  \label{Phi-}
\end{equation}%
If an asymmetric solution exists, taking the difference of Eqs. (\ref{Phi+})
and (\ref{Phi-}) and dividing it by $\Delta \Phi \equiv \Phi _{+}-\Phi
_{-}\neq 0$ yield the following equation:%
\begin{equation}
\left( \sigma _{2}-1\right) \Phi _{+}\Phi _{-}-\left( \Phi _{+}^{2}+\Phi
_{-}^{2}\right) =\eta /\tau .  \label{difference}
\end{equation}

At the \emph{bifurcation point}, where the asymmetric solution splits of
from the symmetric one, $\Phi _{+}=\Phi _{-}\equiv \Phi _{\mathrm{symm}}$,
Eqs. ((\ref{Phi+}), (\ref{Phi-}) and (\ref{difference}) must be satisfied
simultaneously, i.e.,%
\begin{eqnarray}
\eta \Phi _{\mathrm{symm}}+\tau \left( 1+\sigma _{2}\right) \Phi _{\mathrm{%
symm}}^{3} &=&f~\tilde{\delta}(z),  \notag \\
&&  \label{bif} \\
\left( \sigma _{2}-3\right) \Phi _{\mathrm{symm}}^{2} &=&\eta /\tau .  \notag
\end{eqnarray}%
The second equation in system (\ref{bif}) implies that the bifurcation is
only possible if $\sigma _{2}$ is large enough, \textit{viz}., $\sigma _{2}>3
$. Then, the substitution of the second equation into the first one
determines the bifurcation point,
\begin{equation}
\Phi _{\mathrm{symm}}^{\mathrm{(bif)}}=\frac{\left( \sigma _{2}-3\right) }{%
\left( \sigma _{2}-1\right) }\frac{f}{2\eta }\tilde{\delta}(z).  \label{Phi}
\end{equation}%
Further, equating $\Phi _{\mathrm{symm}}^{\mathrm{(bif)}}$, as given by Eq. (%
\ref{Phi}) and by the second equation in system (\ref{bif}), yields the
value of $\tilde{\delta}(z)$ at the bifurcation point:%
\begin{equation}
\left[ f\tilde{\delta}_{\mathrm{bif}}(z)\right] ^{2}=\frac{\left( \sigma
_{2}-1\right) ^{2}}{\left( \sigma _{2}-3\right) ^{3}}\frac{4\eta ^{3}}{\tau
f^{2}}~.  \label{threshold}
\end{equation}%
The meaning of Eq. (\ref{threshold}) is that it gives a \textit{threshold
condition} for the symmetry-breaking bifurcation. Namely, the bifurcation
occurs if the maximum value of $\tilde{\delta}(z)f$, which is its value at
the central point, $z=0$ [for instance, $f\tilde{\delta}(0)=f/\left( \sqrt{%
\pi }\Delta \right) $ in the case of $\tilde{\delta}(z)$ given by Eq. (\ref%
{tilde})] exceeds the threshold value:%
\begin{equation}
\left[ f\tilde{\delta}(0)\right] ^{2}>\left[ f\tilde{\delta}(0)\right] _{%
\mathrm{threshold}}^{2}\equiv \frac{\left( \sigma _{2}-1\right) ^{2}}{\left(
\sigma _{2}-3\right) ^{3}}\frac{4\eta ^{3}}{\tau }~.  \label{SSB}
\end{equation}%
In the present approximation, which corresponds to Eqs. (\ref{Phi+}), (\ref%
{Phi-}), that do not contain derivatives and therefore admit solutions with
discontinuities of the first derivatives, the solution is asymmetric in the
region of $\left\vert z\right\vert <z_{\mathrm{bif}}$, where $z_{\mathrm{bif}%
}$ is to be found from Eq. (\ref{threshold}), and the solution continues as
the symmetric one to $\left\vert z\right\vert \geq z_{\mathrm{bif}}$. This
structure of the solutions resembles phase-separated states in two-component
atomic BEC, found from the respective system of coupled GPEs in the
Thomas-Fermi approximation \cite{Marek}.

\section{Numerical results}

\subsection{Stable symmetric solutions}

The numerical analysis was performed by means of direct simulations of Eqs. (%
\ref{Psi+}), (\ref{Psi-}), using the Crank-Nicolson scheme. Starting with
symmetric initial conditions, the simulations of the equations, where the
regularized delta-function was approximated by Eq. (\ref{tilde}) with
appropriate values of $\Delta $ (in any case, it must be much smaller than
the width of the established state), would always drive the solution to a
stable symmetric localized state. Varying symmetric initial conditions for
fixed parameters of the equations, we always observed the convergence of the
numerical solution to the same stationary state, which is clearly an\textit{%
\ attractor} of the model -- at least, in the class of symmetric initial
conditions (see a discussion of asymmetric configurations below).

\begin{figure}[tbp]
\centering{\includegraphics[width=3in]{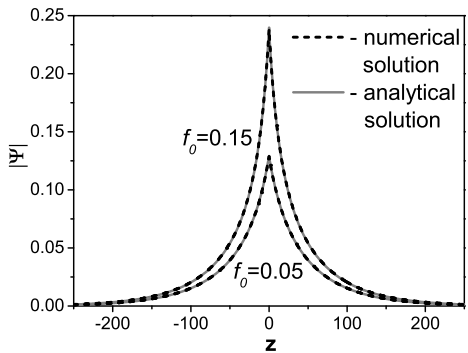}}
\caption{Analytically predicted and numerically generated profiles of the
localized states supported by the $\protect\delta $-function source, at
parameter values $\protect\mu _{0}=0$, $\protect\eta =0.007$, $\protect\tau %
=0.22$, and $\protect\sigma _{1}=\protect\sigma _{2}=2$, $\Delta =1$ for two
different values of the source's strength, $f$ (the small value of $\protect%
\eta $ used here agrees with the condition of the applicability of the
phenomenological model, see the text).}
\label{fig1}
\end{figure}

In all cases when parameter $\delta \mu _{0}$ was chosen as per Eq. (\ref%
{mu0}), so as to support the existence of the analytical symmetric solution
found above, the established state produced by the simulations with $\Delta
\leq 1.5$ virtually exactly coincides with the analytically predicted form
(thus confirming the \emph{stability} analytical solutions). Typical
examples are shown in Fig. \ref{fig1}, for two different values of parameter
$f$ which accounts for the strength of the pumping. Other constants used for
these simulations correspond, after all rescalings, to physical parameters
relevant to the experiment (in particular, $\sigma _{1}=2$ is the value of
the XPM/SPM ratio appropriate to the nonlinear wave interactions mediated by
the cubic terms \cite{Kivshar}).

\begin{figure}[tbp]
\centering{\includegraphics[width=4in]{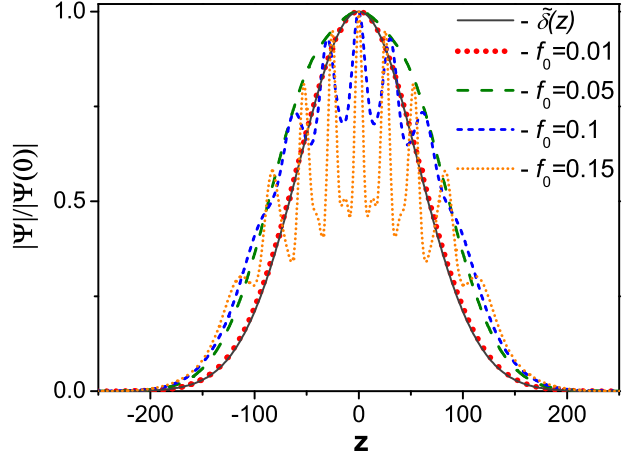}} \caption{(Color
online) Normalized profiles of stable localized states, obtained
with four different values of the strength of the pump, $f$, in
the model with a broad source, corresponding to $\Delta =80$ in
Eq. (\protect\ref{tilde}). Other parameters are the same as in
Fig. \protect\ref{fig1}. Also shown, for comparison, is the shape
of the source $\tilde{\protect\delta}(z)$. The transition from the
single-peak profile to a multi-peak one occurs at some point in
interval $0.05<f<0.10$.} \label{fig2}
\end{figure}

\begin{figure}[tbp]
\centering{\includegraphics[width=3in]{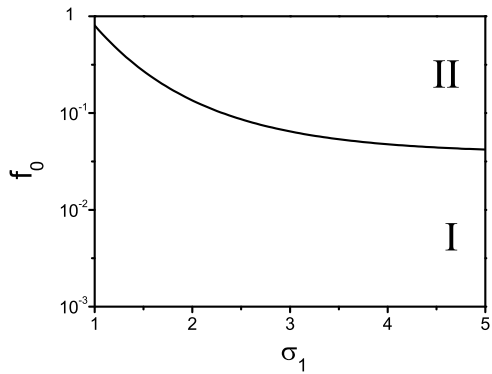}}
\caption{Stationary localized states generated by the simulations
of Eqs. (\protect\ref{Psi+}) and (\protect\ref{Psi-}) feature
single- and multi-peak
shapes in areas I and II, respectively. Other parameters are $\protect\mu %
_{0}=0$, $\protect\eta =0.007$, $\protect\tau =0.22$, and $\protect\sigma %
_{2}=3$.}
\label{fig3}
\end{figure}

Stationary symmetric solutions of another type were produced by
the simulations in the case of a broad source, represented by Eq.
(\ref{tilde}) with much larger values of $\Delta $: the increase
of the source's strength, $f$, and the subsequent growth of the
amplitude of the stationary state leads to a transition from the
simple localized shape with a single maximum to \emph{multi-peak}
patterns, as illustrated in Fig. \ref{fig2}. The crossover between
the single- and multi-peak states in the plane of $\left( \sigma
_{1},f\right) $, which summarizes results of many simulations
performed at different values of the parameters, is displayed in
Fig. \ref{fig3}.

In fact, the transition to the multi-peak patterns with the increase of the
amplitude is a straightforward manifestation of the \textit{modulational
instability }of extended states, which is a property well known both in
conservative models \cite{Kivshar} and in those based on the CGL equations
\cite{CGL}, provided that the nonlinear terms in the model have the sign
corresponding to the self-focusing. As mentioned above, this is indeed the
case for the self-interaction of magnons.

\subsection{Asymmetric solutions}

\begin{figure}[tbp]
\centering{\includegraphics[width=3in]{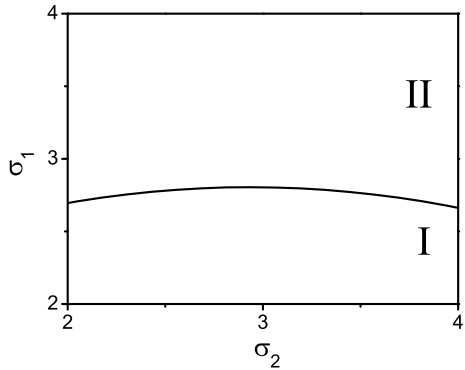}}
\caption{The border between the areas of monostability (I) and bistability
(II), in the plane of parameters $\left( \protect\sigma _{2},\protect\sigma %
_{1}\right) $. Other coefficients are $\protect\mu _{0}=0,$ $\protect\eta %
=0.007,$ $\protect\tau =0.22,$ $f=0.15$, and $\Delta =80$. In area II,
asymmetric initial conditions, with one component equal to zero, generate
persistently asymmetric nonstationary localized solutions, while in area I
the same initial conditions converge to stationary symmetric states. In
either area, symmetric initial conditions always converge to a stationary
symmetric localized state, which may feature a single- or multi-peak shape
in area I, and is always of the multi-peak type in area II.}
\label{fig4}
\end{figure}

The above analysis of the phenomenological model predicted a possibility of
the transition from symmetric to asymmetric states, at least in the strongly
dissipative system. This possibility was systematically tested by running
simulations of Eqs. (\ref{Psi+}) and (\ref{Psi-}) with asymmetric initial
conditions [in fact, this was done by choosing $\Psi _{-}(z,t=0)=0$]. We
have found that, in some region of the parameter space, any initial
condition, symmetric or asymmetric, converges to a single symmetric
localized state. On the other hand, another parameter region exists too,
where the asymmetric initial configuration generates solutions which remain
asymmetric indefinitely long. In fact, this situation implies a \textit{%
bistability }in the system, as the symmetric input would always generate a
stationary symmetric state, at the same values of the parameters. The phase
diagram in the parameter space $\left( \sigma _{2},\sigma _{1}\right) $
showing the regions of the monostability and bistability is displayed in
Fig. \ref{fig4}. It is relevant to stress that, with the variation of the
parameters, the transition from single- to multi-peak symmetric states in
the class of symmetric solutions (see above) always precedes the onset of
the bistability. In other words, in the case of the bistability the
symmetric states coexisting with asymmetric ones always represent multi-peak
patterns.

\begin{figure}[tbp]
\centering{\includegraphics[width=4in]{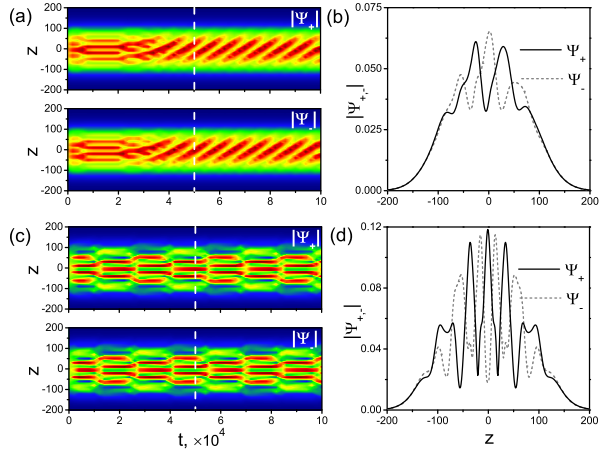}}
\caption{(Color online) Typical examples of the dynamical behavior of
asymmetric states, generated by initial conditions with $\Psi _{-}(z,t=0)=0$
. Parts (a,b) and (c,d) pertain to the pump's strength $f=0.09$ and $0.12$,
respectively. Other parameters are $\protect\mu _{0}=0,$ $\protect\eta =0.007
$, $\protect\tau =0.22$, $\protect\sigma _{1}=\protect\sigma _{2}=3.5$, $%
\Delta =80$ . Panels (a) and (c) display the spatiotemporal evolution of
absolute values of the two fields, $\left\vert \Psi _{\pm }(z,t)\right\vert $%
, while panels (d) and (d) show the profiles of the solutions observed at $%
t=50,000$.}
\label{fig5}
\end{figure}

In fact, the persistently asymmetric states, unlike the symmetric ones,
\emph{never} relax to a stationary shape. While keeping an overall localized
form, they demonstrate quasi-regular oscillations between multi-peaked
configurations in the two components, $\Psi _{+}$ and $\Psi _{-}$. Typical
examples of the dynamical behavior of asymmetric states are displayed in
Fig. \ref{fig5}. It is worthy to note that, at smaller values of the pumping
strength, $f$, each component $\Psi _{\pm }$ of the asymmetric state
generates traveling waves, see Fig. \ref{fig5}(a), while at larger values of
$f$ the nonstationary dynamics is represented by standing waves, as seen in
Fig. \ref{fig5}(c).

In addition to the asymmetry between the components $\Psi _{+}$ and $\Psi
_{-}$, it is relevant to monitor the \emph{spatial} symmetry of these
states. The conclusion is that the solutions are spatially asymmetric at
lower values of the forcing, $0.07<f<0.18,$ and they become spatially
symmetric, with $\Psi _{\pm }(-z)=\Psi _{\pm }(z)$, at higher values of $f,$
although the solution remains nonstationary, with unequal components $\Psi
_{+}$ and $\Psi _{-}$. Pairs of panels (a,c) and (b,d) in Fig. \ref{fig5}
represent, as a matter of fact, typical examples of the spatial asymmetry
and symmetry, respectively (in other words, traveling-wave and standing-wave
patterns, as mentioned above).

\section{Comparison with the experiment}

The model presented above can be used for the description of recent
experiments demonstrating spatiotemporal patterns in magnon condensates
formed through the parametric pumping \cite{threshold}. The respective
pumping field was created by microstrip resonators of two types with
different spatial profiles of the microwave field. The spatiotemporal
evolution of the condensates was studied, using a pumping pulse of a finite
width and applying the technique of the time- and space-resolved Brillouin
light scattering.

Before comparing the theoretical results with experimental data, it is
necessary to stress that the source term in Eqs. (\ref{Psi+}), (\ref{Psi-})
and (\ref{tilde}), $f\tilde{\delta}(z)$, does not directly represent the
pumping field of the resonator used in the experiment. There are two reasons
for that. First, in the experiment the resonator excites pairs of primary
magnons by means of the parametric pumping. The process of parametric
pumping has a threshold, with respect to the pumping field. Thus, although,
as discussed above, the condensate can be created at any value of $f$ which
determines the strength of the source in Eqs. (\ref{Psi+}) and (\ref{Psi-}),
from the experimental point of view the process has a threshold with respect
to the microwave pumping field. Second, the primary magnons create the
condensate as a result of the multi-step thermalization process \cite%
{PRLThermalization}, which is accompanied by the spreading of the
magnon cloud in the physical space. Strictly speaking, to
calculate the spatial profile of the pumping, one needs, first, to
solve the system of equations similar to Eqs. (\ref{Psi+}) and
(\ref{Psi-}), which describe the excitation of the primary
(parametric) magnons. Then, the so generated profiles should be
used in the model accounting for the thermalization process. In
its entire form, this problem can hardly be solved even
numerically. However, since the group velocity of magnons can be
easily found from their spectrum, and the thermalization time is
measured experimentally \cite{Thermalization}, the spreading of
the magnon cloud during the thermalization can be estimated as
being below $100~\mathrm{\mu }$m.

The presence of the effective threshold, can be easily taken into account.
In Ref. \cite{threshold2} it was shown that the number of magnons in the
condensate depends on the strength of the microwave field, $h$, as $\sqrt{%
h^{2}-h_{\mathrm{thr}}^{2}}$. Accordingly, we can represent the source term
as:

\begin{equation}
f\tilde{\delta}(z)\sim \sqrt{h^{2}(z)-(h_{\mathrm{thr}}^{\mathrm{(eff)}})^{2}%
},  \label{thr}
\end{equation}%
where the effect of the spreading of the magnon cloud during the
thermalization process is taken into account by the introduction of the
effective threshold field $h_{\mathrm{thr}}^{\mathrm{(eff)}}$. For values of
$z$ at which $h^{2}(z)-(h_{\mathrm{thr}}^{\mathrm{(eff)}})^{2}$ is negative,
Eq. (\ref{thr}) is replaced by $\tilde{\delta}(z)=0$, since no primary
magnons are excited in this region.

As mentioned above, the parametric pumping excites pairs of primary magnons
with wavenumbers $+k_{0}$ and $-k_{0}$, which makes experimentally relevant
initial conditions for the creation of the condensate symmetric. Therefore,
it is not surprising that only symmetric solutions have been observed in the
experiment.

For the sake of the qualitative comparison, one needs to determine
parameters in Eqs. (\ref{Psi+}) and (\ref{Psi-}). The coefficients
of the linear and nonlinear dissipation, $\eta $\ and $\tau $, can
be extracted from the experimental data via the analysis of the
decay of the condensate when the source (pumping field) is
switched off \cite{Coherence}. For the YIG films used in the
experiments, these coefficients are found to be $\eta \simeq
0.007$ and $\tau \simeq 0.22$ (in the scaled notation,
corresponding to the characteristic nonlinear time of $\tau
_{0}=3.6$ ns). Finally, as mentioned above, due to the fact that
the nonlinearity in the system is caused by the four-magnon
interaction (cubic nonlinearity), the XPM/SPM\ ratios are $\sigma
_{1}=\sigma _{2}=2$.

\begin{figure}[tbp]
\centering{\includegraphics[width=3in]{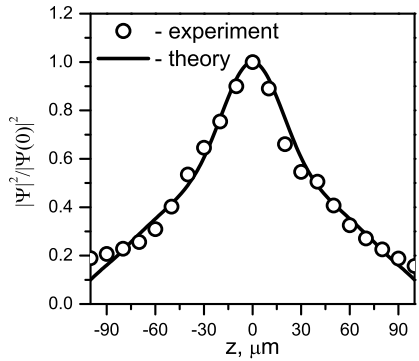}}
\caption{A typical example of the comparison of the normalized stationary
profile of the magnon field, as predicted by the phenomenological equations,
and its experimentally observed counterpart. Further details are given in
the text.}
\label{fig6}
\end{figure}

The comparison between the predictions of the phenomenological
theory and the experimental data borrowed from Ref.
\cite{threshold} is illustrated in Fig. \ref{fig6}, which shows
the spatial profile of the stationary symmetric state calculated
using the above parameters and expression (\ref{thr}) for the
source. The distribution of the parallel component of the
microwave field of the wire resonator was used as $h(z)$, see Fig.
3(a) in Ref. \cite{threshold}. The experimentally measured profile
of the condensate density
is also shown in in Fig. \ref{fig6}. Threshold $h_{\mathrm{thr}}^{\mathrm{%
(eff)}}$ in Eq. (\ref{thr}) was treated as a fitting parameter.

\begin{figure}[tbp]
\centering{\includegraphics[width=3in]{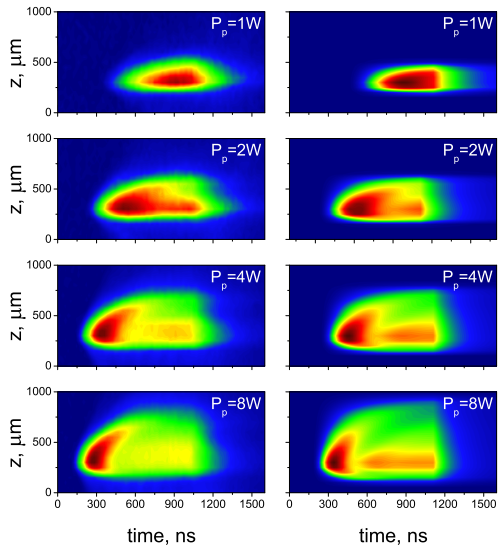}} \caption{(Color
online) \ Spatio-temporal evolution of the density of the magnon
condensate in the case of the pulsed excitation of the microstrip
resonator by the microwave field, for different pumping powers
$P_{\mathrm{p}}$, as indicated in the figure. The left and right
columns display the  experimental and theoretical results,
respectively. Further explanations are given in the text.}
\label{fig7}
\end{figure}

Since Eqs. (\ref{Psi+}) and (\ref{Psi-}) describe the spatiotemporal
dynamics of the condensate, which has been also been explored experimentally
in Ref. \cite{threshold}, it makes sense to compare spatiotemporal patterns
too. Figure \ref{fig7} represents examples of the spatiotemporal evolution
of the magnon field, as predicted by Eqs. (\ref{Psi+}) and (\ref{Psi-}), and
the respective experimental data. The situation presented in the figure
corresponds to the case of the excitation of magnons by the microstrip
resonator \cite{threshold}. Accordingly, the distribution of the microwave
field corresponding to that shown in Fig. 3(b) of \cite{threshold} was used
for the calculation of the source term (\ref{thr}). Point $t=0$ in Fig. \ref%
{fig7} corresponds to the start of the pumping pulse. The delay in the
emergence of the condensate is caused by the process of the thermalization
of magnons, through their nonlinear interactions, as discussed above \cite%
{threshold}. After about 1000 ns, one observes decay of the condensate
density, which is due to the fact that the pumping is switched off at that
time. While calculating the theoretical profiles, we have concluded that the
most appropriate results are produced by the simulations with $\delta \mu
_{0}=-0.011$, which corresponds to $-3\times 10^{6}$ s$^{-1}$, in the
physical units. This value is in agreement with that fact that the pulsed
character of the experiment, with characteristic time $t=1~\mathrm{\mu }$s,
results in the uncertainty of magnon states within a frequency interval $%
\Delta f=2\pi /t\simeq 6.28\times 10^{6}$ s$^{-1}$.

Comparing the theoretical results and experimental data displayed in Fig. %
\ref{fig7}, one can conclude that the present model describes the
spatiotemporal evolution of magnon condensate fairly well. It is clearly
seen that the spatial size of the condensate increases with the growth of
the pumping power. This property is explained by the fact that the spatially
uniform enhancement of the pumping field, just by increasing the microwave
power sent to the resonator, changes the actual profile of the corresponding
source. Indeed, for a larger pumping power, condition $h(z)>h_{\mathrm{thr}%
}^{\mathrm{(eff)}}$, see Eq. (\ref{thr}), is satisfied in a larger interval
of $z$.

\section{Conclusion}

In this work we have proposed a system of semi-phenomenological
one-dimensional equations for the dynamical description of the formation of
spatiotemporal patterns in Bose-Einstein condensates of magnons. The
equations combine essential features of the GPE- and CGL-type models. The
equations include localized source terms, which represent the microwave
field pumping magnons into the condensate. In the limit case of the source
represented by the $\delta $-function, we have found exact analytical
solutions for symmetric localized states of the condensate. The modes also
predicts the possibility of the existence of asymmetric solutions, with
unequal amplitudes of the left- and right-traveling magnon waves. Systematic
simulations of the model have demonstrated that the analytically found
solutions are always stable.

Replacing the $\delta $-function source by the broader one, which is
relevant to the experimental situation, we have found a transition from the
single-peak stationary symmetric solution to stationary multi-peak patterns,
which may be explained as a manifestation of the \ modulational instability
of broad nonlinear modes. While, in direct simultations, symmetric initial
conditions always converge into symmetric stationary states, in a part of
the parameter space asymmetric inputs may generate persistently asymmetric
nonstationary localized modes, which may be realized as patches of traveling
or standing waves, in the cases of low and high amplitudes, respectively.

Comparison with experimental observations demonstrates that the spatial and
spatiotemporal patterns predicted by the phenomenological equations may
provide for a good model of the experiment, with some parameters found from
fundamental characteristics of the ferromagnetic medium, and some others
used for the fitting. Moreover, the model suggests to look for new types of
patterns in the experiment, such as asymmetric ones.

\section*{Acknowledgment}

Support from the Deutsche Forschungsgemeinschaft is gratefully acknowledged.
B.A.M. appreciates hospitality of the Institute for Applied Physics at the
University of M\"{u}nster.

\end{document}